\documentclass[useAMS,usenatbib]{mn2e}
\usepackage{times}
\usepackage{graphics,epsf}
\usepackage{amsmath}                
\usepackage{amsfonts}               
\usepackage{amssymb}                
\usepackage{epsfig}                 
\usepackage{rotating}
\usepackage{color}

\def \s{~\rm{s}}
\def \km{~\rm{km}}

\def \yr{~\rm{yr}}

\def \apj{ApJ}
\def \aap{A\&A}
\def \pasa{PASA}
\def \mnras{MNRAS}
\def \apjl{ApJ Lett.}
\def \apjs{ApJ Suppl. Ser.}
\def \nat{Nature}
\def \actaa{Acta Astron.}
\def \pasp{PASP}

\begin{document}

\title{The explosion of supernova 2011fe in the frame of the core-degenerate scenario}
\author[N. Soker, E. Garc\'\i a--Berro \& L.G. Althaus]
{Noam Soker,$^{1}$ Enrique Garc\'\i a--Berro,$^{2,3}$ and Leandro G. Althaus$^{4}$\\
$^{1}$Department of Physics, Technion -- Israel Institute of Technology, Haifa 32000, Israel; soker@physics.technion.ac.il\\
$^{2}$Departament de F\'\i sica Aplicada, Universitat Polit\`ecnica de Catalunya, c/Esteve Terrades, 5, 08860 Castelldefels, Spain; enrique.garcia-berro@upc.edu\\
$^{3}$Institute for Space Studies of Catalonia, c/Gran Capit\`a 2–4, Edif. Nexus 104, 08034 Barcelona, Spain\\
$^{4}$Facultad de Ciencias Astron\'omicas y Geof\'\i sicas, Universidad Nacional de La Plata, Paseo del Bosque s/n, (1900) La Plata, Argentina; \\althaus@fcaglp.fcaglp.unlp.edu.ar}

\maketitle

\begin{abstract}
We argue that the properties of the Type Ia supernova (SN Ia) SN~2011fe can be best explained within the frame of the core-degenerate (CD) scenario.
In the CD scenario a white dwarf (WD) merges with the core of an asymptotic giant branch (AGB) star and forms a rapidly rotating WD,
with a mass close to and above the critical mass for explosion.
Rapid rotation prevents immediate collapse and/or explosion. Spinning down over a time of $0-10^{10}$~yr brings the WD to explosion.
A very long delayed explosion to post-crystallization phase, which lasts for about $2 \times 10^9 \yr$, leads to the formation of a highly carbon-enriched outer layer.
This can account for the
carbon-rich composition of the fastest-moving ejecta of SN~2011fe.
In reaching the conclusion that the CD scenario best explains the observed properties of SN~2011fe we consider both its specific properties,
like a very compact exploding object and carbon rich composition of the fastest-moving ejecta,
and the general properties of SNe Ia.
\end{abstract}


\section{Introduction}
\label{sec:intro}

SN~2011fe is a typical Type Ia supernova (SN Ia). It was discovered by \cite{Nugentetal2011}, and there is a wealth of observations that constrain its properties.
These constraints can be summarized as follows (\citealt{Chomiuk2013}).
(1) The exploding object had a radius of $R_\ast \la 0.02 R_\odot$ \citep{Bloometal2012}, although other less severe constraints are discussed elsewhere,
e.g., \cite{Mazzalietal2013} who give $R_\ast \la 0.06 R_\odot$, and \cite{PiroNakar2012}.
(2) The fastest-moving ejecta at $v>19\ 400 \km \s^{-1}$ are almost exclusively ($98 \%$ by mass) composed of carbon \citep{Mazzalietal2013}.
(3) The explosion was mildly asymmetric \citep{smithetal2011}.
(4) There are no indications for circumstellar material (CSM).
(5) Very strong constraints on the properties of a possible companion have been placed \citep{Lietal2011}.
Actually, it seems as if the progenitor of SN~2011fe was all alone when it exploded: no binary companion, no material around it,
no violent event much before explosion, and no circumstellar material.
Observations of course only put limits on some physical parameters, but these are so strong that they strongly challenge the double degenerate (DD) scenario
and basically rule out the most popular single degenerate (SD) scenarios for SN~2011fe. There is one speculative SD channel that can better explain SN~2011fe,
which we discuss below.

In this letter we show that these properties can be best explained within the frame of the core-degenerate (CD) scenario.
The plan of the paper is the following. In Sect.~\ref{sec:properties} we confront with observations four basic theoretical scenarios for the formation of the progenitor of SN~2011fe:
($i$) First is the SD scenario (e.g., \citealt{Whelan1973, Nomoto1982, Han2004})
where a WD grows in mass through accretion from a non-degenerate stellar companion.
\cite{Ruiter2011} consider the helium-rich donor scenario \citep{Iben1987} to be a separate category
from the canonical SD scenario. We refer to accretion of helium-rich material under the double-detonation category.
($ii$) Second is the DD scenario \citep{Webbink1984, Iben1984} where two WDs merge after losing angular momentum and energy through the radiation of gravitational waves \citep{Tutukov1979}.
There are suggestions that sub-Chandrasekhar mass remnants can also lead to explosions (e.g., \citealt{vanKerkwijk2010, BadenesMaoz2012}).
($iii$) Third is the Double-Detonation (DDet) mechanism where a sub-Chandrasekhar mass WD accumulates a layer of helium-rich material on the surface,
which under the right conditions can detonate (\citealt{Shenetal2013}, and references therein).
($iv$) Fourth is the CD scenario where a Chandrasekhar or super-Chandrasekhar mass WD is formed at the termination of the common envelope (CE)
phase or during the planetary nebula phase, from a merger of a WD companion with the hot core of a massive asymptotic giant branch (AGB) star
\citep{Livio2003, KashiSoker2011, IlkovSoker2012, IlkovSoker2013, Sokeretal2013, TsebrenkoSoker2013}.
There is some overlap between these scenarios.
For example, a violent merger route of the DD scenario can end up in DDet \citep{Pakmoretal2013}. In Section \ref{sec:carbon}
we discuss how the fastest ejecta of SN~2011fe can be enriched in carbon, as a consequence of carbon-oxygen phase separation upon crystallization.
A short summary is given in Section \ref{sec:summary}.

\section{The properties of SN~2011fe}
\label{sec:properties}

\cite{Chomiuk2013} presents a general summary of the properties of SN~2011fe and the way they constrain the SD and DD scenarios. Here we limit the discussion to some
specific properties that hold the key to rank the likelihood of the different scenarios.
We also briefly discuss the strong and weak points of each scenario in relation to general properties of SN Ia.
Moreover, since PTF~11kx is frequently mentioned as being the result of a SD event, we would like to express here our stand that this cannot be the case,
because the massive CSM of PTF~11kx can be much better explained by the CD scenario \citep{Sokeretal2013}.

As SN~2011fe appears to be a normal SN Ia, we consider only scenarios that are claimed to account for a large fraction of SNe Ia.
The WD-WD collision model \citep{KatzDong2012, Kushniretal2013} can account for at most few per cent of all SNe Ia \citep{Hamersetal2013, Prodanetal2013},
and is not discussed here.
One can reach this conclusion by considering the different demands on this process. These include a small fraction of
triple systems \citep{LeighGeller2013}, the requirement that progenitor of SN Ia have $M \la 1.7 M_\odot$ to be compatible with the delay-time distribution (\citealt{Greggioetal2008}),
and the limitation that merger cannot take place before the formation of two WDs \citep{Hamersetal2013}.

We now turn to discuss some specific items of the scenarios. Most awkward to the SD scenario is that no companions are found in nearby supernovae remnants (SNRs) of SN Ia.
This holds for SN~2011fe as well. The second general weak point of the SD scenario is that it can account neither for the shape of the time-delay distribution
(DTD) nor for the total number of SN Ia (e.g., \citealt{Nelemansetal2013}). The strongest prediction of the SD scenario is the presence of hydrogen in the CSM.
However, when a hydrogen-rich CSM is detected it is too massive to be accounted for by the SD scenario, as for PTF~11kx \citep{Sokeretal2013}.
One way to overcome some of the problems of the SD scenario is to assume that rotation of the WD delays the explosion till long after accretion ceases
\citep{Justham2011, DiStefanoetal2011}. This delay has some common properties with the delay of the CD scenario,
and can explain some properties of SN~2011fe much as the CD scenario does. However, this does not solve other problems of the SD scenario,
such as that it is expected to explain only a small fraction of SNe Ia, and that it is not clear that the WD can grow by accretion to the Chandrasekhar mass limit.

The strongest character of the DD scenario is that it well explains the DTD (e.g., \citealt{MaozMannucci2012}; \citealt{Ruiteretal2013}; \citealt{Nelemansetal2013}).
However, the ignition process and whether  sub-Chandrasekhar systems can explode are still open questions.
Merges of two WDs might release large amounts of gravitational energy in the form of electromagnetic radiation. If the WD-WD merge occurs much before the explosion,
we would expect to see many transient events with luminosity not much below, and even higher, than in SN Ia. These are not observed.
This and other considerations led to the study of prompt ignition mechanisms, such as the violent merge scenario \citep{Pakmoretal2011, Pakmoretal2012, Pakmoretal2013}
that was confronted with SN~2011fe by \cite{Ropkeetal2012}. Violent merges lead to highly asymmetrical explosions \citep{Pakmoretal2011, Pakmoretal2012, Pakmoretal2013},
with a large departure from axisymmetry.
\cite{smithetal2011} conducted a spectropolarimetry study of SN~2011fe and concluded that ``[the small polarization]
is suggestive that there is some small amount of global asymmetry in the ejecta of SN~2011fe, perhaps even suggesting axial symmetry in the event.''
This is compatible with the finding that well resolved close by SN Ia remnants are close to being spherical \citep{Lopezetal2011}.
The close to spherical morphologies pose a strong challenge to the violent merge ignition mechanism.

In the studies of \cite{Pakmoretal2011, Pakmoretal2012} carbon is ignited on the accreting WD. This can reduce the carbon abundance in the fastest moving gas,
as in the violent merger model for subluminous SN Ia studied by \cite{Pakmoretal2011}. This is contrary to the observations of SN~2011fe that has $98 \%$
carbon-rich material in the fastest-moving ejecta \citep{Mazzalietal2013}. The carbon-rich fastest ejecta is also problematic for models based on helium accretion,
such as the DDet scenario \citep{Mazzalietal2013}. Instead, \cite{Mazzalietal2013} prefer accretion of hydrogen that is burned to carbon during the explosion.
In the CD scenario the long time that laps between the core-WD merge and the explosion allows carbon to separate from oxygen when the WD crystallizes.
We elaborate on this in Section \ref{sec:carbon}.

In the violent merger process only part of the mass lost by the destructed WD is accreted on to the more compact WD. The rest of the mass expands up to a
distance of about $0.03-0.04 R_\odot$ from the exploding WD \citep{Pakmoretal2011, Pakmoretal2012, Pakmoretal2013}. In the simulations of \cite{Pakmoretal2012}
the shock breaks out of the gas at a radius of about $0.04-0.05 R_\odot$. This is on the edge of what can be compatible with the limits on the size of the exploding
WD in SN~2011fe. The actual limits on the violent merger process and on any RLOF process by the size of the exploding WD of SN~2011fe are even more tight.
The reason is that the mass transfer proceeds via an accretion disk lasting for at least several tens of orbital periods, greater than about $100 \s$.
It is very likely that a disk-wind and/or jets are blown during this period with velocities close to the escape velocity from the accreting WD,
about $5000 \km \s^{-1}$ \citep{Ji2013}. Therefore, outflowing gas perpendicular to the equatorial plane will reside at distances of about
$1 R_\odot$ at the time of explosion.

The DDet scenario has been thoroughly discussed in recent years (e.g., \citealt{Simetal2012, Shenetal2013}), as it can well account for ignition,
as well as for other properties of SN Ia (e.g. \citealt{Ruiter2011}). The helium can be supplied from a degenerate or a non-degenerate companion.
Observational constraints on the DDet depend on the nature of the mass donor star. If it is non-degenerate (degenerate) then some of the drawbacks
of the SD scenario (DD scenario) are applicable. In addition, for the specific case of SN~2011fe the carbon-rich fastest ejecta is difficult to
explain with helium accretion \citep{Mazzalietal2013}.

The conclusion of this discussion is that there is no scenario exempt of problems. As for the specific case of SN~2011fe, it seems that the CD scenario does the best.
The most puzzling observation that the fastest ejecta is $98 \%$ rich carbon is dealt with in the next section.

\section{The carbon-rich fast ejecta}
\label{sec:carbon}

As previously discussed, none of the standard scenarios is able to satisfactorily account for the presence of an almost pure carbon shell in the fastest,
outermost region of the ejecta. The CD scenario, however, is able to explain this feature.
It is expected that in the merge of the an AGB star and a WD the tiny H envelope (about $10^{-4}\, M_\odot$) and He buffer (about $10^{-2}\, M_\odot$)
are ejected or burned as a consequence of the dynamical interaction \citep{Danetal2013}.
Thus, we foresee that the result of such interaction is a WD with a bare carbon-oxygen core.
Actually, there is observational evidence for WDs devoid of these external H and He-rich layers \citep{Gansicke2010}.
Moreover, SPH simulations \citep{L-A2009} show that the resulting WD has a rapidly rotating and hot convective corona,
which is prone to the  magnetorotational instability (\citealt{G-B2012}; on the MRI see \citealt{BalbusHawley1991}).
Consequently, the remnant of the merge should be a rapidly rotating, magnetized white dwarf.
The outer hydrogen and helium layers of the core and WD will actually carry the extra angular momentum and be expelled from the merged product.
If the delay time is sufficiently long, as we propose for SN~2011fe, the ejecta of the merge has long gone and the WD goes through crystallization.
In passing we note that in the SD and DDet scenarios accretion from a non-degenerate companion keeps the core warm and prevents crystallization.
During the crystallization phase the concentrations of carbon and oxygen are not the same in the liquid and the solid phase \citep{Nature1,Nature2}.
The oxygen abundance is higher in the solid phase.
Hence, the denser oxygen-rich solid sinks and the carbon-rich liquid is homogeneized by Rayleigh-Taylor instabilities \citep{Isern1,Isern2}.
The result of this process is that, as crystallization proceeds, the outer layers of the WD become richer in carbon.

\begin{figure}
\includegraphics[width=0.9\columnwidth,clip]{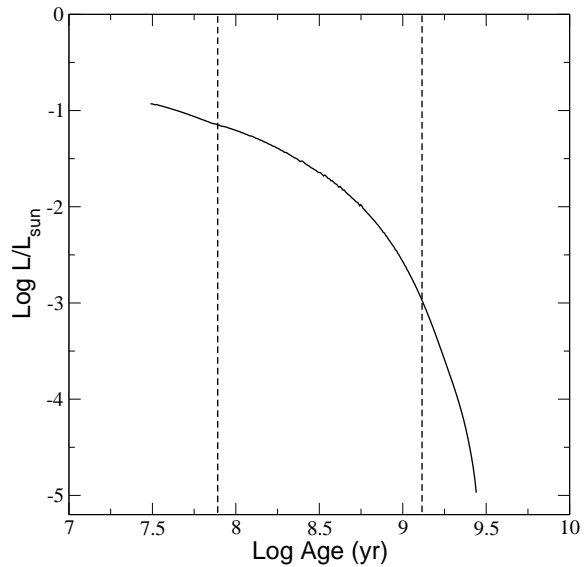}
\caption{Cooling sequence of the bare nucleus of a $1.38\, M_\odot$ carbon-oxygen WD. The leftmost dashed vertical line shows when crystallization starts at the centre of the star,
while the rightmost marks when the core is 95\% crystallized.}
\label{fig1}
\end{figure}

\begin{figure}
\includegraphics[width=0.9\columnwidth,clip]{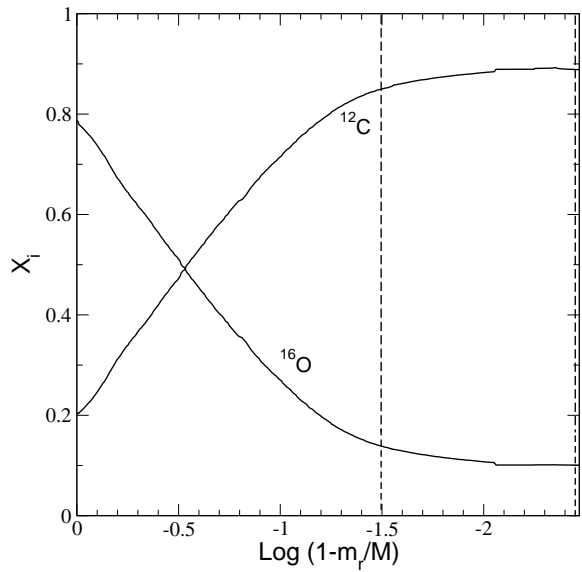}
\caption{Chemical profile the WD of Fig.~1 when the core is 95\% crystallized. Shown is the composition from the center of the WD to
$10^{-3} M$ from its surface as a function of $\log (1-m_r/M)$. Here $m_r$ is the mass inner to radius $r$ and $M$ is the total mass of the WD.
The vertical dashed line shows the mass coordinate at
which the carbon mass abundance is largely enhanced by crystallization $(X_{\rm C}\ga 0.85)$.}
\label{fig2}
\end{figure}

To check whether or not this is a viable scenario to explain the enhanced carbon abundance of the very outer layers of SN~2011fe we followed the evolution of the bare core of a
$1.38\, M_\odot$ WD from the knee in the Hertzsprung-Russell diagram until the luminosity of the WD was as low as $\log(L/L_\odot)\simeq -5.0$ (see \citealt{Renedo} for details).
The knee in the HR diagram is the phase when the luminosity and temperature of the young WD start to decrease.
We considered that the WD had no He nor H outer layers since, as mentioned, these layers are very likely ejected during the merger. The results are shown in Figs~\ref{fig1} and~\ref{fig2}.
The core is 95\%  crystallized at $t_{\rm cool}\simeq 1.4$~Gyr, a relatively short delay (Fig. \ref{fig1}). By that time the luminosity of the WD is $\log(L/L_\odot)\simeq -3.1$,
its effective temperature is $\log T_{\rm eff}\simeq 4.3$, and it has an outer layer of mass $\Delta M \approx 0.045 \, M_\odot$
where the carbon mass abundance is $X_{\rm C}\simeq 0.9$ (Fig. \ref{fig2}). For lower luminosities the carbon abundance in the very outer layers does not change appreciably.
Thus, we expect that, after a sufficiently long time, the explosion of such WD results in the very outer layers being largely enhanced in carbon,
in good, but not perfect, agreement with the observations of SN~2011fe.

\section{Summary}
\label{sec:summary}

SN~2011fe is an archetypical type Ia supernova that was observed shortly after it exploded. The early detection allowed us to strongly constrain the properties of its progenitor.
Studies in the past two years confronted some theoretical models with these constraints. However, the core degenerate (CD) scenario was not considered in any of these studies.
Here we argue that the CD scenario best accounts for the properties of SN~2011fe.
Our arguments, which were discussed in Section~\ref{sec:properties}, are
compared to the most relevant properties of SN~2011fe, as collected by \cite{Chomiuk2013}, in Table~\ref{tab:Table1}. From our presentation of the observed and expected properties
of SN~2011fe it is evident that no scenario is free of problems.
However, it seems that the CD scenario best survives the different limits on the properties of the SN~2011fe progenitor.
\begin{table*}
\begin{center}
  \caption{Confronting four SN Ia scenarios with the properties of SN~2011fe.}
    \begin{tabular}{| p{2.7cm} | p{2.7cm}| p{2.7cm}| p{2.7cm}| p{2.7cm} |}
\hline  
{}                       & {Single Degenerate}    & {Double Degenerate} & {Double Detonation} & {Core Degenerate} \\
\hline  
{SN~2011fe: $R_\ast<0.02 R_\odot${$^{[1]}$}}  &  {\textcolor[rgb]{0.00,0.59,0.00}{Expected}} & {\textcolor[rgb]{0.98,0.00,0.00}{Marginal for violent merger} }
                         & {Depends on donor type  }    & {\textcolor[rgb]{0.00,0.59,0.00}{Expected}}  \\
\hline  
{SN~2011fe: $98 \%$ Carbon in fastest ejecta}  & {Possible$^{[2]}$ } & {\textcolor[rgb]{0.98,0.00,0.00}{Not expected}}
                                    &   {\textcolor[rgb]{0.98,0.00,0.00}{Problematic}{$^{[2]}$}}  & {\textcolor[rgb]{0.00,0.59,0.00}{Separation after crystallization}}\\
\hline  
{SN~2011fe: Mildly asymmetric explosion}       &   {\textcolor[rgb]{0.00,0.59,0.00}{Expected}} & \textcolor[rgb]{0.98,0.00,0.00}{{Highly asymmetric explosion}}
                                      &{Depends on mass transfer process  } & {\textcolor[rgb]{0.00,0.59,0.00}{Expected}}\\
\hline  
{SN~2011fe: No Circumstellar material}       &   {\textcolor[rgb]{0.98,0.00,0.00}{Magic is needed to hide companion}} & {\textcolor[rgb]{0.00,0.59,0.00}{Expected}}
                                &{Depends on mass transfer process  } & {\textcolor[rgb]{0.00,0.59,0.00}{Expected in most cases}}\\ \cline{1-1} \cline{3-5}
{SN~2011fe: Strong limits on a companion}       &  {\textcolor[rgb]{0.98,0.00,0.00}{ and its wind}} & {\textcolor[rgb]{0.00,0.59,0.00}{Expected}}
                                       & {Depends on donor} & {\textcolor[rgb]{0.00,0.59,0.00}{Expected}} \\
\hline  
{General: Strong characteristics}       &   {\textcolor[rgb]{0.00,0.59,0.00}{Accreting massive WDs exist}} & {\textcolor[rgb]{0.00,0.59,0.00}{Explains very well the delay time distribution (DTD)}}
                                        & {\textcolor[rgb]{0.00,0.59,0.00}{Ignition easily achieved}}
                                        & {\textcolor[rgb]{0.00,0.59,0.00}{Explains both SN Ia with H-CSM and symmetric explosion}} \\
\hline  
{General: Weak characteristics}    & {\textcolor[rgb]{0.98,0.00,0.00}{(1) Cannot account for DTD; (2) Companions not found}}
                                   & {\textcolor[rgb]{0.98,0.00,0.00}{Ignition process (violent merger has too-asymmetric ejecta)}}
                                 & {\textcolor[rgb]{0.98,0.00,0.00}{Same as for SD and DD, depending on type of donor}}
                                 & {\textcolor[rgb]{0.98,0.00,0.00}{More work on (1) delay-parameter; (2) Merge during CE; (3) Find massive single WDs}}  \\
\hline  
{PTF~11kx: Hydrogen rich and massive CSM}  & {\textcolor[rgb]{0.98,0.00,0.00}{CSM too massive}} & {\textcolor[rgb]{0.98,0.00,0.00}{Not expected at all}}
                                  & {\textcolor[rgb]{0.98,0.00,0.00}{CSM too massive}} & {\textcolor[rgb]{0.00,0.59,0.00}{{Expected in rare cases}}{$^{[3]}$}} \\
\hline  
     \end{tabular}
  \label{tab:Table1}\\
\end{center}
\begin{flushleft}
\small Notes:\\
\small \footnotemark[1]{$R_\ast$ is the radius of the exploded star.} \\
\small \footnotemark[2]{\cite{Mazzalietal2013}}\\
\small \footnotemark[3]{\cite{Sokeretal2013}}
\end{flushleft}
\end{table*}

Although the CD scenario does well with the constraints on the progenitor of SN~2011fe, some of the properties of the CD scenario are still poorly determined and deserve further study.
\begin{enumerate}
\item Carbon enriched ejecta (Section~\ref{sec:carbon}). The post-crystallization model of WDs presented in Section~\ref{sec:carbon} brings the carbon enrichment to about $90 \%$,
a little short of the observed $98 \%$. We envisage two possibilities that can improve the agreement with observations, although we do not discard other effects.
First, the derived abundances are based on the spectral fittings of SN~2011fe with the W7 model of \cite{Iwamoto99} and an improved model (W7$^+$)
specifically designed to obtain a better fit to the observed spectrum. These models have very different density profiles, such that a small change in the
slope of the density profile of the ejecta may change the carbon abundance by a few percent, bringing our results in better agreement with observations.
Secondly, the carbon abundance in the very outer layers depends on the initial carbon abundance in the inner core, which depends on the
$^{12}$C$(\alpha,\gamma)^{16}$O reaction rate, as well as on the temperature and density profiles of the progenitor star.
This reaction rate is still uncertain, and small changes in the cross section may result in an enhanced carbon abundance in the core (e.g., \citealt{Salaris1997}).
Nevertheless, an in depth study of these effects should be made in subsequent works.
\item The core-WD merge process. In this process either the core or the WD are destroyed and accreted by the other object.
When the accreting object approaches the Chandrasekhar limit it contracts and releases gravitational energy.
We speculate that this regulates the final merged product to be of
about the Chandrasekhar mass \citep{TornambePiersanti2013}, with some mass spreading due to rapid rotation.
\item The delay parameter. To account for the delay time distribution of ${dN_{\rm Ia}}/{d t_{\rm SF}} \propto t^{-1}$ \citep{MaozMannucci2012},
where $t_{\rm SF}$ is the time since star formation,
a parameter to which the delay time is very sensitive is required. Namely, the time from star formation to explosion sensitively depends on some parameter $\aleph$,
such that $\tau_{\rm e} \propto \aleph ^{\eta}$ with $\eta \gg 1$. For the CD scenario this can be the angular momentum loss \citep{IlkovSoker2012, TornambePiersanti2013}
or the decay of the magnetic field, or another parameter. This will be studied in a future work.
\item The properties of the merged product should be determined in order to search for such massive WDs.
\cite{Tout2008} considered a merge of a WD with a core of an AGB star to explain the formation of massive rotating WDs with strong magnetic fields, and
\cite{Wickramasinghe2000} commented that such WDs might be more likely to form SN Ia. WDs with strong magnetic fields and mass around the Chandrasekhar mass
are predicted to exist by our scenario. However, their observational properties should be better determined. For example, whether the merge process removes all hydrogen
and even helium.
\end{enumerate}

\section*{Acknowledgments}
We thank the referee, Christopher Tout, for a very detail and helpful report.
This  research  was partially  supported  by  AGAUR, by  MCINN  grants
AYA2011--23102 and AYA08-1839/ESP, by  the European Union FEDER funds, and by the ESF EUROGENESIS project (grant EUI2009-04167).


\label{lastpage}


\begin{thebibliography}{}

\bibitem[Althaus et al.(2010)]{Renedo} Althaus, L.~G.,
Garc{\'{\i}}a-Berro, E., Renedo, I., et al.\ 2010, \apj, 719, 612

\bibitem[Badenes \& Maoz(2012)]{BadenesMaoz2012} Badenes, C., \& Maoz, D.\ 2012, \apjl, 749, L11

\bibitem[Balbus \& Hawley(1991)]{BalbusHawley1991} Balbus, S.~A., \& Hawley, J.~F.\ 1991, \apj, 376, 214

\bibitem[Bloom et al.(2012)]{Bloometal2012} Bloom, J.~S., Kasen, D., Shen, K.~J., et al.\ 2012, \apjl, 744, L17

\bibitem[Chomiuk(2013)]{Chomiuk2013} Chomiuk, L.\ 2013,  PASA, in press, arXiv:1307.2721

\bibitem[Dan et al.(2013)]{Danetal2013} Dan, M., Rosswog, S., Brueggen, M., \& Podsiadlowski, P.\ 2013,  arXiv:1308.1667

\bibitem[Di Stefano et al.(2011)]{DiStefanoetal2011} Di Stefano, R., Voss, R., \& Claeys, J.~S.~W.\ 2011, \apjl, 738, L1

\bibitem[G{\"a}nsicke et al.(2010)]{Gansicke2010} G{\"a}nsicke, B.~T., Koester, D., Girven, J., Marsh, T.~R., \& Steeghs, D.\ 2010, Science, 327, 188

\bibitem[Garc\'\i a-Berro et al.(1988)]{Nature1} Garc\'\i a-Berro, E., Hernanz, M., Isern, J., \& Mochkovitch, R.\ 1988, \nat, 333, 642

\bibitem[Garc{\'{\i}}a-Berro et al.(2010)]{Nature2} Garc{\'{\i}}a-Berro, E., Torres, S., Althaus, L.~G., et al.\ 2010, \nat, 465, 194

\bibitem[Garc{\'{\i}}a-Berro et al.(2012)]{G-B2012} Garc{\'{\i}}a-Berro, E., Lor{\'e}n-Aguilar, P., Aznar-Sigu{\'a}n, G., et al.\ 2012, \apj, 749, 25

\bibitem[Greggio et al.(2008)]{Greggioetal2008} Greggio, L., Renzini, A., \& Daddi, E.\ 2008, \mnras, 388, 829

\bibitem[Hamers et al.(2013)]{Hamersetal2013} Hamers, A.~S., Pols, O.~R., Claeys, J.~S.~W., \& Nelemans, G.\ 2013, \mnras, 430, 2262

\bibitem[Han \& Podsiadlowski(2004)]{Han2004} Han, Z., \& Podsiadlowski, P.\ 2004, \mnras, 350, 1301

\bibitem[Iben et al.(1987)]{Iben1987} {{{ {Iben, I., Jr., Nomoto, K., Tornambe, A., \& Tutukov, A.~V.\ 1987, \apj, 317, 717} }}}

\bibitem[Iben \& Tutukov(1984)]{Iben1984} Iben, I., Jr., \& Tutukov, A.~V.\ 1984, \apjs, 54, 335

\bibitem[Ilkov \& Soker(2012)]{IlkovSoker2012} Ilkov, M., \& Soker, N.\ 2012, \mnras, 419, 1695

\bibitem[Ilkov \& Soker(2013)]{IlkovSoker2013} Ilkov, M., \& Soker, N.\ 2013, \mnras, 428, 579

\bibitem[Isern et al.(1997)]{Isern1} Isern, J., Mochkovitch, R., Garc\'\i a-Berro, E., \& Hernanz, M.\ 1997, \apj, 485, 308

\bibitem[Isern et al.(2000)]{Isern2} Isern, J., Garc{\'{\i}}a-Berro, E., Hernanz, M., \& Chabrier, G.\ 2000, \apj, 528, 397

\bibitem[Iwamoto et al.(1999)]{Iwamoto99} Iwamoto, K., Brachwitz,
F., Nomoto, K., et al.\ 1999, \apjs, 125, 439

\bibitem[Ji et al.(2013)]{Ji2013} Ji, S., Fisher, R.~T., Garc{\'{\i}}a-Berro, E., et al.\ 2013, \apj, 773, 136

\bibitem[Justham(2011)]{Justham2011} Justham, S.\ 2011, \apjl, 730, L34

\bibitem[Kashi \& Soker(2011)]{KashiSoker2011} Kashi, A., \& Soker, N.\ 2011, \mnras, 417, 1466

\bibitem[Katz \& Dong(2012)]{KatzDong2012} Katz, B., \& Dong, S.\ 2012, arXiv:1211.4584

\bibitem[Kushnir et al.(2013)]{Kushniretal2013} Kushnir, D., Katz, B., Dong, S., Livne, E., \& Fern{\'a}ndez, R.\ 2013, arXiv:1303.1180

\bibitem[Leigh \& Geller(2013)]{LeighGeller2013} Leigh, N.~W.~C., \& Geller, A.~M.\ 2013, \mnras, 432, 2474

\bibitem[Li et al.(2011)]{Lietal2011} Li, W., Bloom, J.~S., Podsiadlowski, P., et al.\ 2011, \nat, 480, 348

\bibitem[Livio \& Riess(2003)]{Livio2003} Livio, M., \& Riess, A.~G.\ 2003, \apjl, 594, L93

\bibitem[Lopez et al.(2011)]{Lopezetal2011} Lopez, L.~A., Ramirez-Ruiz, E., Huppenkothen, D., Badenes, C., \& Pooley, D.~A.\ 2011, \apj, 732, 114

\bibitem[Lor{\'e}n-Aguilar et al.(2009)]{L-A2009} Lor{\'e}n-Aguilar, P., Isern, J., \& Garc{\'{\i}}a-Berro, E.\ 2009, \aap, 500, 1193

\bibitem[Maoz \& Mannucci(2012)]{MaozMannucci2012} Maoz, D., \& Mannucci, F.\ 2012, \pasa, 29, 447

\bibitem[Mazzali et al.(2013)]{Mazzalietal2013} Mazzali, P., Sullivan, M., Hachinger, S., et al.\ 2013, arXiv:1305.2356

\bibitem[Nelemans et al.(2013)]{Nelemansetal2013} Nelemans, G., Toonen, S., \& Bours, M.\ 2013, IAU Symposium, 281, 225 arXiv:1204.2960

\bibitem[Nomoto(1982)]{Nomoto1982} Nomoto, K.\ 1982, \apj, 253, 798

\bibitem[Nugent et al.(2011)]{Nugentetal2011} Nugent, P.~E., Sullivan, M., Cenko, S.~B., et al.\ 2011, \nat, 480, 344

\bibitem[Pakmor et al.(2011)]{Pakmoretal2011} Pakmor, R., Hachinger, S., R{\"o}pke, F.~K., \& Hillebrandt, W.\ 2011, \aap, 528, A117 

\bibitem[Pakmor et al.(2012)]{Pakmoretal2012} Pakmor, R., Kromer, M., Taubenberger, S.,  Sim, S. A., R"opke, F. K., \& Hillebrandt, W.\ 2012, \apjl, 747, L10 

\bibitem[Pakmor et al.(2013)]{Pakmoretal2013} Pakmor, R., Kromer, M., Taubenberger, S., \& Springel, V.\ 2013, \apjl, 770, L8

\bibitem[Piro \& Nakar(2012)]{PiroNakar2012} Piro, A.~L., \& Nakar, E.\ 2012, arXiv:1211.6438

\bibitem[Prodan et al.(2013)]{Prodanetal2013} Prodan, S., Murray, N., \& Thompson, T.~A.\ 2013, arXiv:1305.2191

\bibitem[R{\"o}pke et al.(2012)]{Ropkeetal2012} R{\"o}pke, F.~K., Kromer, M., Seitenzahl, I.~R., et al.\ 2012, \apjl, 750, L19 

\bibitem[Ruiter et al.(2011)]{Ruiter2011} Ruiter, A.~J., Belczynski, K., Sim, S.~A., Hillebrandt, W., Fryer, C. L., Fink, M., \& Kromer, M.\ 2011, \mnras, 417, 408

\bibitem[Ruiter et al.(2013)]{Ruiteretal2013} Ruiter, A.~J., Sim, S.~A., Pakmor, R., et al.\ 2013, \mnras, 429, 1425

\bibitem[Salaris et al.(1997)]{Salaris1997} Salaris, M., Dominguez,
I., Garcia-Berro, E., et al.\ 1997, \apj, 486, 413

\bibitem[Shen et al.(2013)]{Shenetal2013} Shen, K.~J., Guillochon, J., \& Foley, R.~J.\ 2013, \apjl, 770, L35

\bibitem[Sim et al.(2012)]{Simetal2012} Sim, S.~A., Fink, M., Kromer, M., et al.\ 2012, \mnras, 420, 3003

\bibitem[Smith et al.(2011b)]{smithetal2011} Smith, P.~S., Williams, G.~G., Smith, N.,  Milne, P.~A., Jannuzi, B.~T., \& Green, E.~M.\ 2011b, arXiv:1111.6626  

\bibitem[Soker et al.(2013)]{Sokeretal2013} Soker, N., Kashi, A., Garc\'ia-Berro E., Torres, S., \& Camacho, J.\ 2013, \mnras, 431, 1541

\bibitem[Tornamb{\'e} \& Piersanti(2013)]{TornambePiersanti2013} Tornamb{\'e}, A., \& Piersanti, L.\ 2013, \mnras, 431, 1812

\bibitem[Tout et al.(2008)]{Tout2008} Tout, C.~A., Wickramasinghe, D.~T., Liebert, J., Ferrario, L.,
      \& Pringle, J.~E.\ 2008, \mnras, 387, 897

\bibitem[Tsebrenko \& Soker(2013)]{TsebrenkoSoker2013} Tsebrenko, D., \& Soker, N.\ 2013, arXiv:1305.1845

\bibitem[Tutukov \& Yungelson(1979)]{Tutukov1979} Tutukov, A.~V., \& Yungelson, L.~R.\ 1979, \actaa, 29, 665

\bibitem[van Kerkwijk et al.(2010)]{vanKerkwijk2010} van Kerkwijk, M.~H., Chang, P., \& Justham, S.\ 2010, \apjl, 722, L157

\bibitem[Webbink(1984)]{Webbink1984} Webbink, R.~F.\ 1984, \apj, 277, 355

\bibitem[Whelan \& Iben(1973)]{Whelan1973} Whelan, J., \& Iben, I., Jr.\ 1973, \apj, 186, 1007

\bibitem[Wickramasinghe \& Ferrario(2000)]{Wickramasinghe2000} Wickramasinghe, D.~T.,     \& Ferrario, L.\ 2000, \pasp, 112, 873

\end{thebibliography}
\end{document}